\documentstyle[preprint,revtex]{aps}

\begin{document}
\draft

\begin{title}
Magnetic excitations and effects of magnetic fields \\
on the spin-Peierls transition in CuGeO$_3$
\end{title}

\author{Jos\'e Riera and Sergio Koval}

\begin{instit}
Instituto de F\'{\i}sica Rosario, Consejo Nacional de Investigaciones
Cient\'{\i}ficas y T\'ecnicas,  \\
and Departamento de F\'{\i}sica,
Universidad Nacional de Rosario,    \\
Avenida 27 de Febrero 210b,  2000-Rosario, Argentina.
\end{instit}

\receipt{}

\begin{abstract}
We analyze the magnetic excitations of a spin-1/2 antiferromagnetic
Heisenberg model with alternating nearest neighbor interactions and
uniform second neighbor interactions recently proposed
to describe the spin-Peierls transition in CuGeO$_3$.
We show that there is good agreement between the calculated
excitation dispersion relation and the experimental one.
We have also shown that this model reproduces satisfactorily the
experimental results for the magnetization vs. magnetic field curve
and its saturation value. The model proposed also reproduces
qualitatively some features of the magnetic phase diagram of this
compound and the overall behavior of the magnetic specific heat
in the presence of applied magnetic fields.
\end{abstract}


\newpage

Interest in low-dimensional quantum spin systems has been intensified
in recent years by the observation of a rich variety of behaviors in
new materials and by important theoretical
results.\cite{haldane,ladders}
In particular, the spin-Peierls (SP) transition is one of the most
attractive phenomena and it was first discovered and studied in
organic compounds.\cite{bray83}
At the SP transition, the high-temperature phase with uniform
spin-$1/2$ Heisenberg antiferromagnetic chains (``U phase") changes
into the low-temperature phase with dimerized or alternating chains
(``D phase"). In the dimerized phase the system is characterized by
the presence of an energy gap (the ``SP gap") between the singlet
ground state and the first triplet state.

Recently the SP transition was observed in the inorganic material
CuGeO$_3$ by Hase, Terasaki and Ichinokura.\cite{hase1}
They showed that the magnetic susceptibility in CuGeO$_3$ along all
directions drop drastically at the transition temperature
$\rm T_{SP} \approx 14\:K$.
The SP nature of this transition has been confirmed by several
experimental results.\cite{hase,lorenzo,harris,nishi,brill,pouget}
The quasi one-dimensional (1D) nature of this compound is given by
the presence of weakly coupled linear Cu$^{2+}$ spin-1/2 chains
along the $c$ direction.
The transition is the result of the coupling between these chains
and the three-dimensional phonons.\cite{lorenzo,harris}

In order to characterize the SP phase in this compound, it is
important to investigate its thermodynamic properties in the
presence of magnetic fields because as it is well-known in organic
SP materials they have characteristic behaviors below $\rm T_{SP}$
and large magnetic fields.\cite{bonner87}
The phase diagram of SP systems in the magnetic field and temperature
plane, as determined experimentally from specific heat and ac
susceptibility measurements, shows the uniform and dimerized phases
at low magnetic fields, and another magnetic phase (``M phase") at
high magnetic fields.
In the absence of magnetic fields,
the dimerization of the chains corresponds to a lattice distortion
with wave vector $\pi$. For any $\rm T < T_{SP}$, the magnetic field
above a critical value $\rm H_c(T)$ favors a distortion at some
other wave vector or another type of distortion. The M phase could
then correspond to another conmensurate, discommensurate (or magnetic
soliton) or inconmensurate phase. In the case of CuGeO$_3$, this M
phase has not yet been fully characterized.

Recently, in Ref. \cite{riera} (``RD" in the following) we attempted
a description of the SP transition in CuGeO$_3$ with a spin-1/2 1D
Heisenberg antiferromagnetic model coupled to the three-dimensional
phonons in the adiabatic approximation.
The spin part of this model is defined by the Hamiltonian:
\begin{eqnarray}
\rm H_s = \sum_{i} \rm J_i {\bf S}_i \cdot {\bf S}_{i+1} +
\alpha \rm J \sum_{i} {\bf S}_i \cdot {\bf S}_{i+2}
\label{hamj1j2}
\end{eqnarray}
\noindent
where the index i runs over the lattice sites (i=1,...,N, N:
number of sites) with periodic boundary conditions. The exchange
coupling constant $\rm J_i = J (1 + \delta (-1)^i)$ where
$\delta = \delta(\rm T)$ is a temperature dependent dimensionless
quantity which gives the degree of dimerization.
The model Eq. (\ref{hamj1j2}) has been studied so far in the limits
$\alpha = 0$\cite{bonner83}, and $\delta = 0$.\cite{okamoto}

The parameters of the model were chosen so as to give a good fit of
the experimental magnetic susceptibility and to reproduce the
measured excitation gap at $\rm T=0$.\cite{harris,nishi} The
resulting parameters are: $\rm J=160\ K$, $\alpha=0.36$
and $\delta(\rm T=0)=0.014$.
As it is well-known, the 1D Heisenberg antiferromagnetic
model with first and second neighbor interactions has a finite
singlet-triplet gap
for $\alpha > \alpha_c \approx 0.2412$.\cite{okamoto,castilla}
Thus, the value of $\alpha=0.36$ implies that there is a finite
gap even in the absence of dimerization. This
spin gap was estimated in RD to be $\approx 0.015$ in units of
J, or $2.4\ \rm K$, much smaller than the smallest measured
value for CuGeO$_3$.\cite{harris}
In a more recent study Castilla, Chakravarty and
Emery\cite{castilla} argued, from an analysis of the neutron
scattering measurements of the superlattice peaks\cite{harris},
that the value of $\alpha$ should be below the critical one.
For this reason, they proposed the following set of
parameters: $\rm J=150\ K$, $\alpha=0.24$, and
$\delta(\rm T=0)=0.030$.
However, we think that this argument, based on the theory
of Cross and Fisher\cite{crossfish} which uses a
Luther-Peschel-type treatment of the spin Hamiltonian,
is not conclusive.
Moreover, we should notice that there is a rather large interchain
coupling along the $b$ direction, $\rm J_b/J \approx 0.1$\cite{nishi}
that could lead to a different effective $\alpha_c$.

The purpose of this work is to describe using the model proposed in
RD other features of the SP phase in CuGeO$_3$, in particular the
nature of magnetic excitations and the behavior of various
thermodynamic properties in the presence of an external magnetic
field as have been revealed by an increasing amount of experimental
results.
We want to show also that the model and numerical procedure developed
in RD could provide an alternative complete theory of the
spin-Peierls transition and that it could be considered as an useful
tool to analyze experimental data.
These studies will be performed by using exact diagonalization
techniques on finite chains.
In these calculations we adopt the parameters given in RD
which lead to a somewhat better fit to the experimental magnetic
susceptibility as well as for the dispersion relation as we shall
see below than the parameter set given in Ref. \cite{castilla}.
Most of the calculations were performed also for this latter
parameter set and we shall discuss the main qualitative differences
found.

We start our study with an analysis of magnetic excitations
at zero temperature and in the absence of a magnetic field.
For this study we have employed the Lanczos algorithm for lattices
of up to 24 sites.
We have also calculated the dynamical magnetic structure factor
$\rm S(k,\omega)$
for all the wave vectors $\rm k$.\cite{haasdago}
This dynamical response has the largest intensity at $\rm k=0$
(corresponding to $\pi$ in the uniform lattice) and
hence we shall concentrate our study to this momentum. The results
for $\rm S(\rm k~=~0,\omega)$ are depicted in Fig. 1a for
$\alpha = 0.36$ and $\delta = 0.014$. We also show the results for
the nondimerized model for comparison.
The first peak corresponds to the singlet-triplet excitation.
This peak has the largest weight, in agreement with experimental
results.\cite{nishi}
We also note other excitations at higher frequencies and
with lower intensities. It is interesting to notice that the second
peak in the nondimerized case is split by a finite $\delta$.
The relative intensity of this second peak with respect to the
first one remains approximately constant as N is increased.
Similar results are obtained with the set of parameters of
Ref. \cite{castilla}, although the second peak is located at
slightly higher frequencies than for our set of parameters.
In order to understand the origin of these excitations, let us
examine them as a function of $\rm k$.
The results for the $\rm N=24$ chain are shown in Fig. 1b together
with recent neutron scattering results.\cite{nishi}
It can be seen that the parameter set of RD fits slightly better
the experimental data than the parameter set of
Ref. \cite{castilla}.
In Fig. 1c we show the excitation energy of various states
as a function of k for the same chain. At $\rm k=0$ it is convenient
to label each energy level by $\rm E_i(S,P)$ where S is the total
spin, P ($=+1,-1$) is the quantum number associated to the reflection
symmetry, and $\rm i=0 (1)$ for the ground state (first excited
state).
For $\rm N = 4n$, n integer, the singlet-triplet spin gap is
$\Delta_{st}=\rm E_0(1,-)-E_0(0,+)$. For $\rm N = 4n+2$, the values
of P are reversed. The values of P quoted below correspond to the
case $\rm N = 4n$.

By studying the excitation spectra for several lattice sizes, we
obtained that the position of the second peak in
$S(\rm k~=~0,\omega)$ is $\omega_1=\rm E_1(1,-)-E_0(0,+)$.
The excited singlet state $\rm E_1(0,+)$ appears inside the
singlet-triplet gap but this is just a finite
size effect as can be seen in Fig.1d.
This feature is absent for $\alpha=0.24$, i.e.
$\rm E_1(0,+) > E_0(1,-)$ for any lattice size.
In Fig. 1d, we plot $\Delta_{st}$, $\omega_1$ and
$\omega_2=\rm E_1(0,+)-E_0(0,+)$ as a function of $\rm N^{-1}$.
The data do not seem to vary linearly with $\rm N^{-2}$ and there is
some spread in the extrapolated values depending on the fitting
function chosen.
In Fig. 1d tentative extrapolations using the function
$\omega = \omega_{\infty} + \rm a N^{-1} + \rm b N^{-2}$ are shown.
It is apparent that in the bulk limit the singlet-triplet
excitation becomes the lowest excitation in agreement with the case
$\alpha = 0.24$ and with experimental results.
Besides, several trial extrapolations indicate that there would
remain a finite gap between the first and second peaks of
$\rm S(\rm k~=~0,\omega)$ in the thermodynamic limit for
$\alpha = 0.36$.
The results are less conclusive for $\alpha = 0.24$.
Finally, we want to notice that the first excited triplet state
has $\rm P=-1$ for $\alpha = 0.36$ but $\rm P=+1$ for
$\alpha = 0.24$.

We turn now to the study of thermodynamic properties in the presence
of an applied magnetic field H along the $c$ direction.
In order to study the SP phase at finite temperature and in the
presence of a magnetic field we minimize the free energy for the
total Hamiltonian with respect to $\delta$ to obtain
$\delta_{eq}(\rm H,T)$.
The total Hamiltonian of the system consists of the spin part
given by Eq. (\ref{hamj1j2}) with the addition of an elastic term
$\sim \delta^2$ and the Zeeman term $\rm g \mu_0 S^z H$, where g is
the Land\'e factor (we take $\rm g=2$), $\mu_0$ is the Bohr magneton
and $\rm S^z$ is the z-component of the total spin.
The dimerization parameter $\delta_{eq}(\rm H,T)$ is, for a given T,
a decreasing function of H (Fig. 2a). Consequently, the spin gap
and the critical temperature $\rm T_{SP}(H)$ are also suppressed by
the application of an external magnetic field, as discussed in
several theoretical studies.\cite{bonner87,crossfish,bulaevsk}
Once $\delta_{eq}(\rm H,T)$ is known, the computation of the
thermodynamic properties we are interested in is straightforward.
This procedure only requires the computation of the eigenvalues of
the Hamiltonian (\ref{hamj1j2}) which is readily done by
using the Householder algorithm for chains of up to 14 sites.

Recently the magnetization curve was measured in the presence of
ultra-high magnetic fields up to $\approx \rm 500\ T$.\cite{nojiri}
The measurement of the magnetization and its saturation can be
used to determine the values of the antiferromagnetic exchange
interactions for an adopted model Hamiltonian.
Nojiri {\it et al}\cite{nojiri} noticed that the
experimental magnetization curve shows good agreement with the
theoretical curve obtained for the
spin-1/2 uniform Heisenberg antiferromagnetic chain with
nearest-neighbor coupling $\rm J = 183\ K$.
However, as discussed in RD, this model does not satisfactorily
reproduce the magnetic susceptibility data.
We plot in Fig. 2b the magnetization M versus applied magnetic
field H obtained with the model proposed in RD for $\rm N=14$ site
chain at $\rm T=5$ and $\rm 8\ K$.
We also include the experimental results corresponding to $\rm T=6$
and $\rm 10\ K$.\cite{nojiri}
Our numerical curves present the typical steplike structure due to
the finite lattices involved.
We observe an overall good agreement of the theoretical curve
compared to the experiment results.
The temperature dependence of our results is somewhat weaker than
the experimental one, specially for high magnetic fields where
presumably the three-dimensional effects are more important.
We point out here that without further
parameters to adjust in our model, we have also obtained the
value of the saturation field ($\approx \rm 253\ T$) in
reasonable agrement with the experimental result.
We obtain very similar results with the parameters of
Ref. \cite{castilla}.

The knowledge of $\delta_{eq}(\rm H,T)$ enables us to discuss some
features of the magnetic phase diagram of CuGeO$_3$.
This magnetic phase diagram has been partially determined
experimentally by measuring the magnetization in magnetic fields up
to $25\ \rm T$,\cite{hase} and it presents the three phases
characteristic of a SP system. Hase {\it et al.} \cite{hase} showed
that as the applied magnetic field is increased, at temperatures
below $\rm T_{SP}(0)$, there is a nonlinear increase of
the magnetization above a certain $\rm H_c(T)$.\cite{ohta}
This nonlinear increase, together with the presence of hysteresis,
indicates that there is a first order transition between the
dimerized phase and the high-field M phase for $\rm T \leq 10\ K$.
The hysteresis disappears above this temperature and the D-U phase
boundary becomes a second order one.
This boundary is determined in our microscopic theory by the
condition $\delta_{eq}(\rm H,T)=0$.
The results for $\rm N=12, 14$ and $16$ site chains\cite{lanczos}
using our parameters set are plotted in Fig. 3.
We also show for comparison the analytical results obtained by
Cross\cite{cross} using the theory developed in
Ref. \cite{crossfish}.
The most important assumption of this theory is that
$\rm T_{SP}(0) \ll J$, moderately satisfied in this compound.
For low magnetic fields, our results closely agree with the curve
calculated by Cross corresponding to the D-U transition (dashed
line) and with experimental results.\cite{hase,hori}

For higher magnetic fields, the theory of Ref. \cite{cross} predicts
a transition between the high-field magnetic (assumed incommensurate)
and the uniform phases, which is also shown in Fig. 3 (full line).
There are still no experimental results for CuGeO$_3$ corresponding to
this M-U transition.
In order to study the incommensurability of the M phase and the
transitions D-M and M-U, the nearest neighbor coupling constants in
our model should be generalized as $\rm J_i = J (1 + \delta cos(qi))$
where $\rm q= q(H)$ is the component of the phonon wave vector
parallel to the chain direction.
The dimerization studied so far corresponds to the commensurate
value $\rm q=\pi$. In the presence of a magnetic field
the value of q is the one that minimizes the free energy of the
system. At present there are important numerical limitations
to accomplish such a program. Moreover, it should be noticed that
the interchain exchange interactions have stronger effects in the
M phase.\cite{inagaki}
Although we have not attempted an extrapolation to the bulk limit,
the curves obtained with our model in the high-field region should
converge as the chain size is increased to an upper bound to the D-M
boundary.
The determination of the first-order nature of the D-M phase
transition is also out of the scope of this work because of the
limitation in the numerically accessible chain sizes.

Finally, we study the magnetic specific heat $\rm C_m$ in the
presence of a magnetic field.
The results for $\rm C_m/T$ obtained for the $\rm N=14$ site chain
as a function of $\rm T/T_{SP}$ for
$\rm H = 0$, 5 and $\rm 10\ T$ are shown in Fig. 4.
Experimentally, it has been found that the specific heat shows a
sharp anomaly at $\rm T_{SP}$
due to the SP transition\cite{kuroe,oseroff}. This peak is clearly
seen in our results.
Consistently with the results shown in Fig. 2b, the position of the
peak corresponding to the SP transition decreases as the
magnetic field is increased and the low temperature
part of the curve moves upward.
These changes of $\rm C_m/T$ with magnetic field are also in
qualitative agreement with the experimental data from
Ref. \cite{oseroff}. The differences in shape between the theoretical
and experimental curves are presumably also a finite size effect.

In summary, we have analyzed magnetic excitations of the spin-1/2
antiferromagnetic Heisenberg model with alternating nearest neighbor
interactions and uniform second neighbor interactions proposed in
RD to describe the spin-Peierls transition in CuGeO$_3$. These
results are among the first reported for this very complicated model
and they could be relevant for future neutron scattering experiments.
There is good agreement between the calculated excitation
dispersion relation and the experimental one.
We have also shown that this model reproduces satisfactorily the
experimental results for the magnetization vs. magnetic field curve
and its saturation value. The model proposed also reproduces
qualitatively some features of the magnetic phase diagram of this
compound and the overall behavior of the magnetic specific heat
in the presence of applied magnetic fields.

We acknowledge H. Nojiri for sending us their results for
M vs. H prior publication.
S. K. acknowledge helpful discussions with A. Dobry and
A. Greco during the early stages of this work.

\newpage

\figure{a) Spin dynamical structure factor $\rm S(k=0,\omega)$ for
$\rm N=20$ with $\alpha=0.36$.
The solid (dashed) line corresponds to $\delta=0.014$ $(\delta=0)$.
b) Excitation dispersion for the $\rm N=24$ chain. The solid
line corresponds to $\rm J=160\ K$, $\alpha=0.36$ and $\delta=0.014$
and the dashed line to $\rm J=150\ K$, $\alpha=0.24$ and
$\delta=0.030$. The experimental data (open circles) is taken from
Ref. \cite{nishi}.
c) Excitation dispersion of various states for the $\rm N=24$ chain
with $\rm J=160\ K$, $\alpha=0.36$ and $\delta=0.014$.
The two lowest singlet (triplet) states are indicated with
squares (circles).
d) $\Delta_{st}$ (solid circles), $\omega_1$ (squares) and $\omega_2$
(open circles), defined in the text, versus $\rm N^{-1}$.
The dotted lines are quadratic fits in $\rm N^{-1}$.
\label{fig1}}

\figure{a) The reduced dimerization constant $\delta \rm (H)$
versus the applied magnetic field H at $\rm T=3\ K$
(circles), $\rm 5\ K$ (squares), $\rm 7\ K$ (diamonds) and
$\rm 9\ K$ (triangles).
b) The magnetization M as a function of magnetic field at
different temperatures.
The theoretical results for $\rm N=14$ are plotted with lines
and the experimental data from Ref. \cite{nojiri} with symbols.
\label{fig2}}

\figure{Magnetic phase diagram of the SP system CuGeO$_3$.
Open circles, solid squares and open diamonds correspond to
$\delta_{eq} \rm (T,H)=0$ for N=12, 14 and 16 sites
respectively. We also show
the experimental data from Ref. \cite{hase} (crosses). The
results obtained by Cross\cite{cross} are plotted with
solid and dashed lines.
\label{fig3}}

\figure{Magnetic specific heat as a function of temperature,
for different values of the applied magnetic field H.
\label{fig4}}


\newpage


\begin{references}

\bibitem{haldane} F. D. M. Haldane, Phys. Rev. Lett. {\bf 50}, 1153
       (1983).

\bibitem{ladders} E. Dagotto, J. Riera, and D. J. Scalapino, Phys.
       Rev. B {\bf 45}, 5744 (1992).

\bibitem{bray83} For a review, see J. W. Bray, L. V. Interrante, I.
      S. Jacobs, and J. C. Bonner, in {\em Extended linear
      chain compounds}, edited by J. S. Miller, (Plenun, New York
      1983), Vol 3, pp 353-415.

\bibitem{hase1} M. Hase, I. Terasaki and K. Uchinokura, Phys. Rev.
       Lett. {\bf 70}, 3651 (1993).

\bibitem{hase} M. Hase {\it et al.}, Phys. Rev. B {\bf 48}, 9616
       (1993).

\bibitem{lorenzo} J. E. Lorenzo {\it et al.}, Phys. Rev. B {\bf 50},
       1278 (1994); K. Hirota {\it et al.}, Phys. Rev. Lett.
       {\bf 73}, 736 (1994).

\bibitem{harris} Q. J. Harris {\it et al.}, Phys. Rev. B {\bf 50},
       12606 (1994).

\bibitem{nishi} M. Nishi, O. Fujita and J. Akimitsu, Phys. Rev. B
       {\bf 50} 6508 (1994).

\bibitem{brill} T. M. Brill {\it et al.}, Phys. Rev. Lett. {\bf 73},
       1545 (1994).

\bibitem{pouget} J. P. Pouget {\it et al.}, Phys. Rev. Lett.
       {\bf 72}, 4037 (1994).

\bibitem{bonner87} J. C. Bonner, J. A. Northby, I. S. Jacobs and L.
         V. Interrante, Phys. Rev. {\bf B} 35, 1791 (1987).

\bibitem{riera} J. Riera and A. Dobry, Phys. Rev. B {\bf 51}, 16098
       (1995).

\bibitem{bonner83} J. C. Bonner {\it et al.}, Phys. Rev. B {\bf 27},
      248 (1983); M. Azzouz and C. Bourbonnais, preprint (1995).

\bibitem{okamoto} K. Okamoto and K. Nomura, Phys. Lett. A {\bf 169},
       433 (1992); and references therein.

\bibitem{castilla} G. Castilla, S. Chakravarty, and V. J. Emery,
       preprint (1995).

\bibitem{crossfish} M. C. Cross and D. S. Fisher, Phys. Rev. B
       {\bf 19}, 402 (1979).

\bibitem{haasdago} A study of $\rm S(k,\omega)$ for all $\rm k$ and
       using the parameters of Ref. \cite{castilla} was recently
       reported by S. Haas and E. Dagotto, preprint (1995).

\bibitem{bulaevsk} L. N. Bulaevskii, A. I. Buzdin and D. I. Khomski,
      Solid State Comm. {\bf 27}, 5 (1978).

\bibitem{nojiri} H. Nojiri {\it et al.}, preprint (1995).

\bibitem{hori} H. Hori {\it et al.}, J. Phys. Soc. Jpn. {\bf 63}, 18
     (1994).

\bibitem{lanczos} The data corresponding to $\rm N= 16$ are the
       result of an approximate calculation using the Lanczos
       algorithm. It is assumed that the use of translational
       symmetries breaks the level degeneracy.

\bibitem{ohta} See also H.Ohta {\it et al.}, J. Phys. Soc. Jpn.
      {\bf 63}, 2870 (1994).

\bibitem{cross} M. C. Cross, Phys. Rev. B {\bf 20}, 4606 (1979).

\bibitem{inagaki} S. Inagaki and H. Fukuyama, J. Phys. Soc. Jpn.
       {\bf 52}, 877 (1983).

\bibitem{kuroe} H. Kuroe {\it et al.}, J. Phys. Soc. Jpn. {\bf 63},
        365 (1994); S. Sahling {\it et al.}, Solid State Comm.
        {\bf 92}, 423 (1994).

\bibitem{oseroff} S. B. Oseroff {\it et al.}, Phys. Rev. Lett.
       {\bf 74}, 1450 (1995).

\end{references}
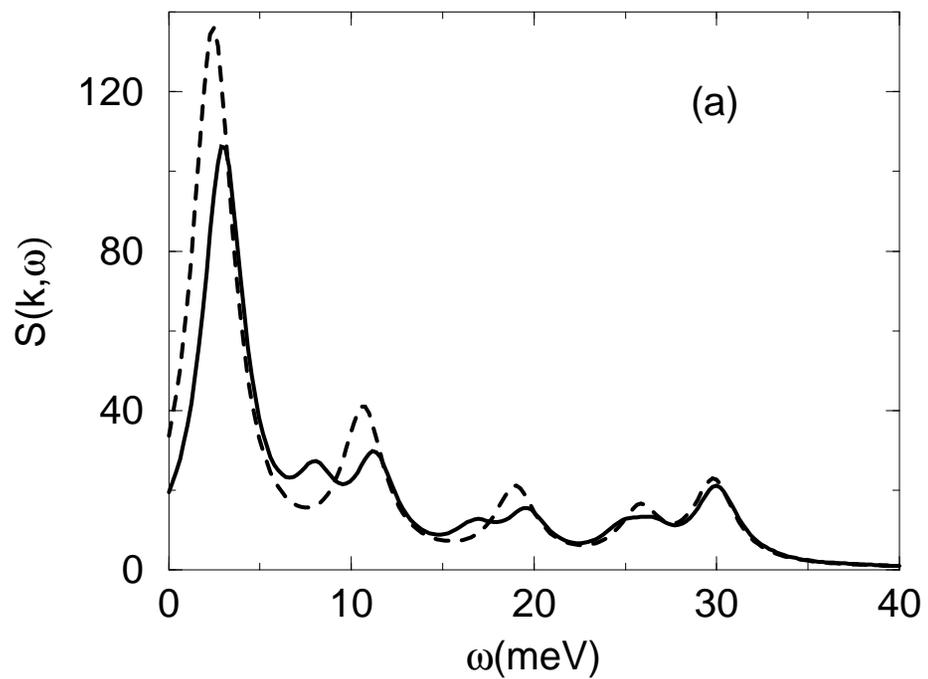
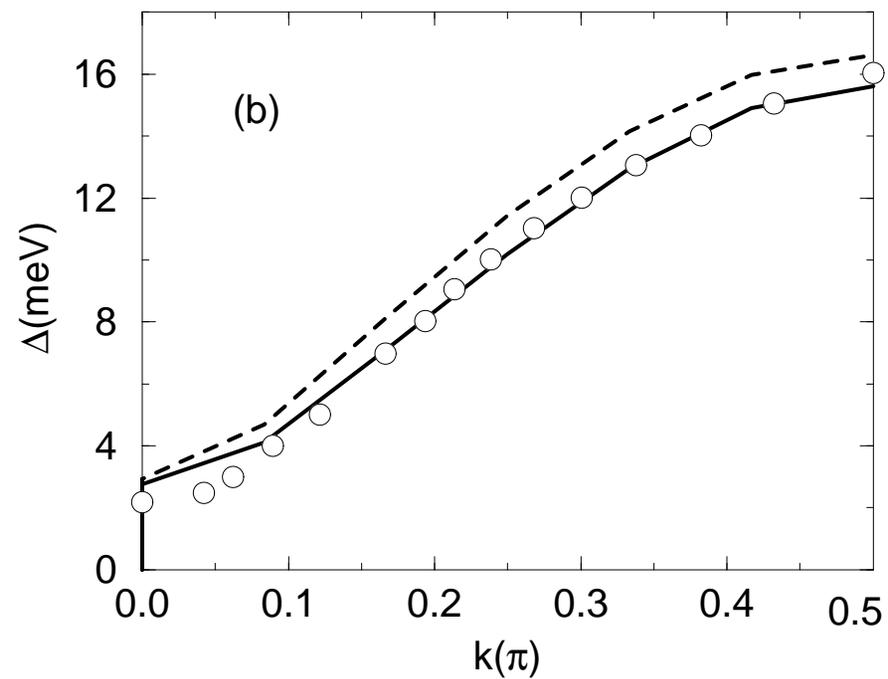
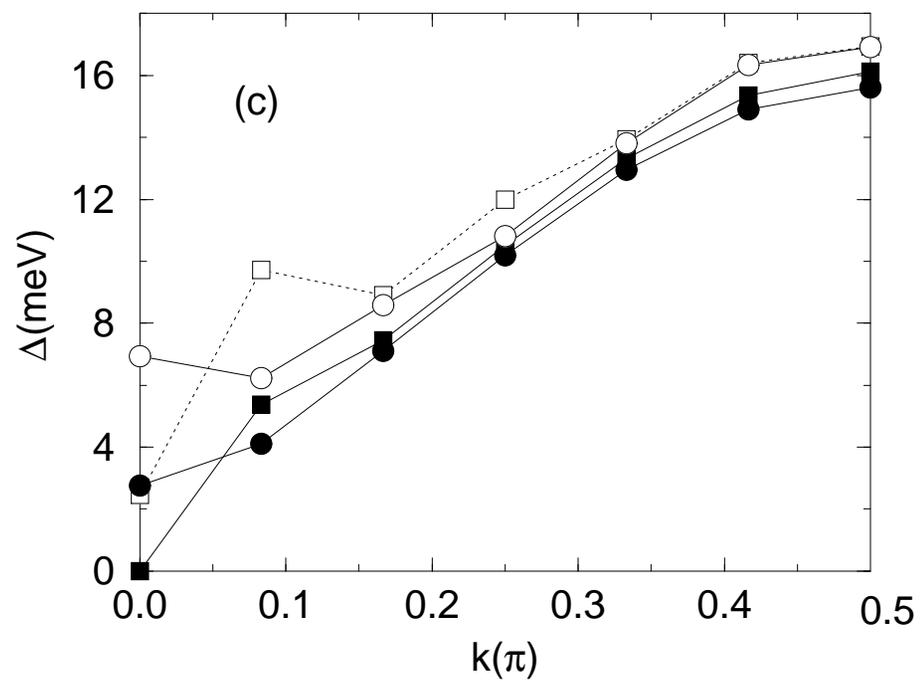
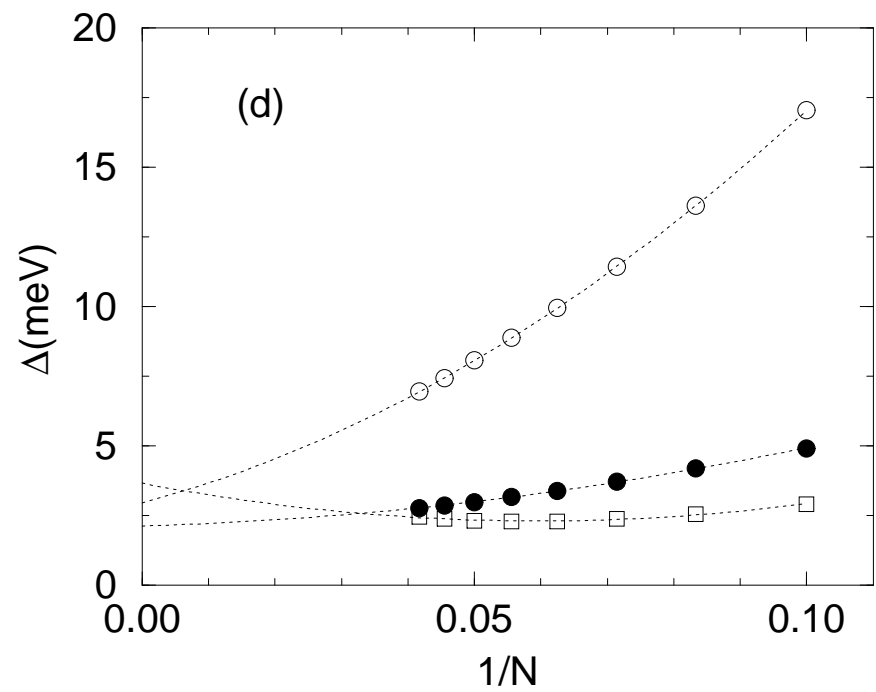
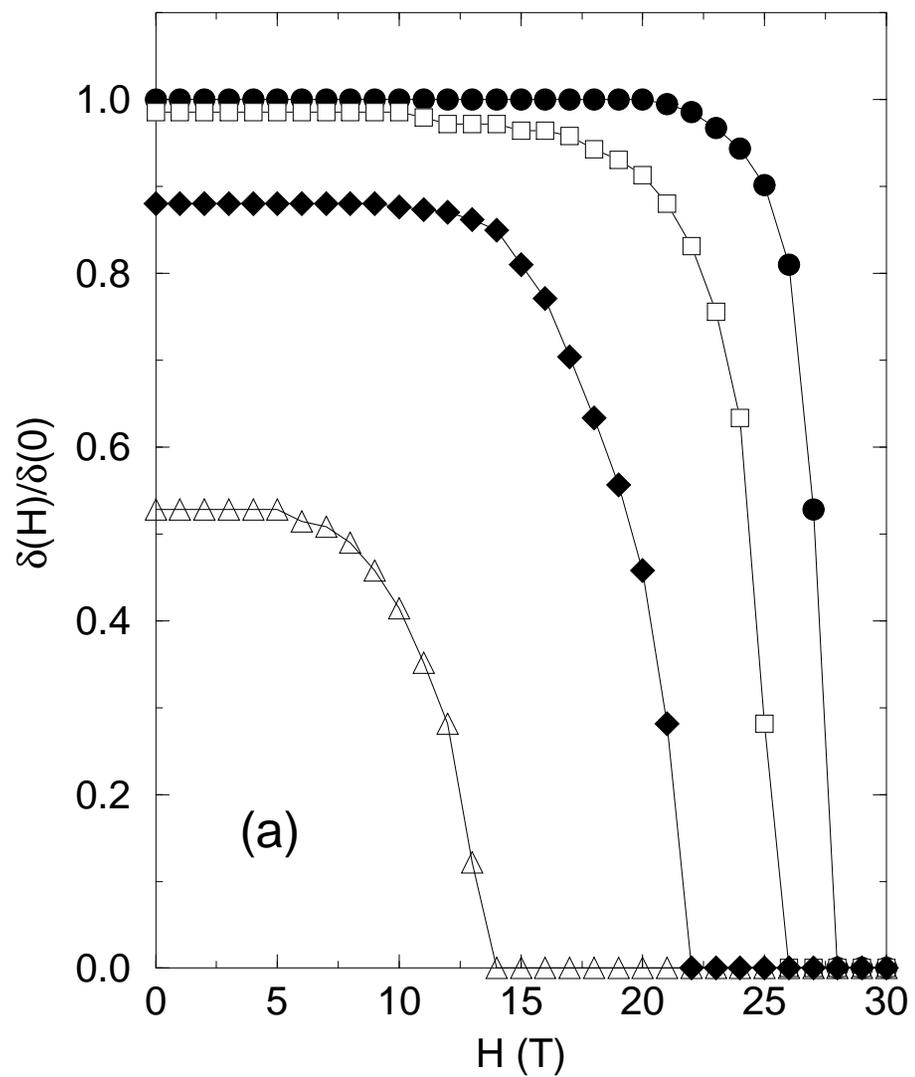
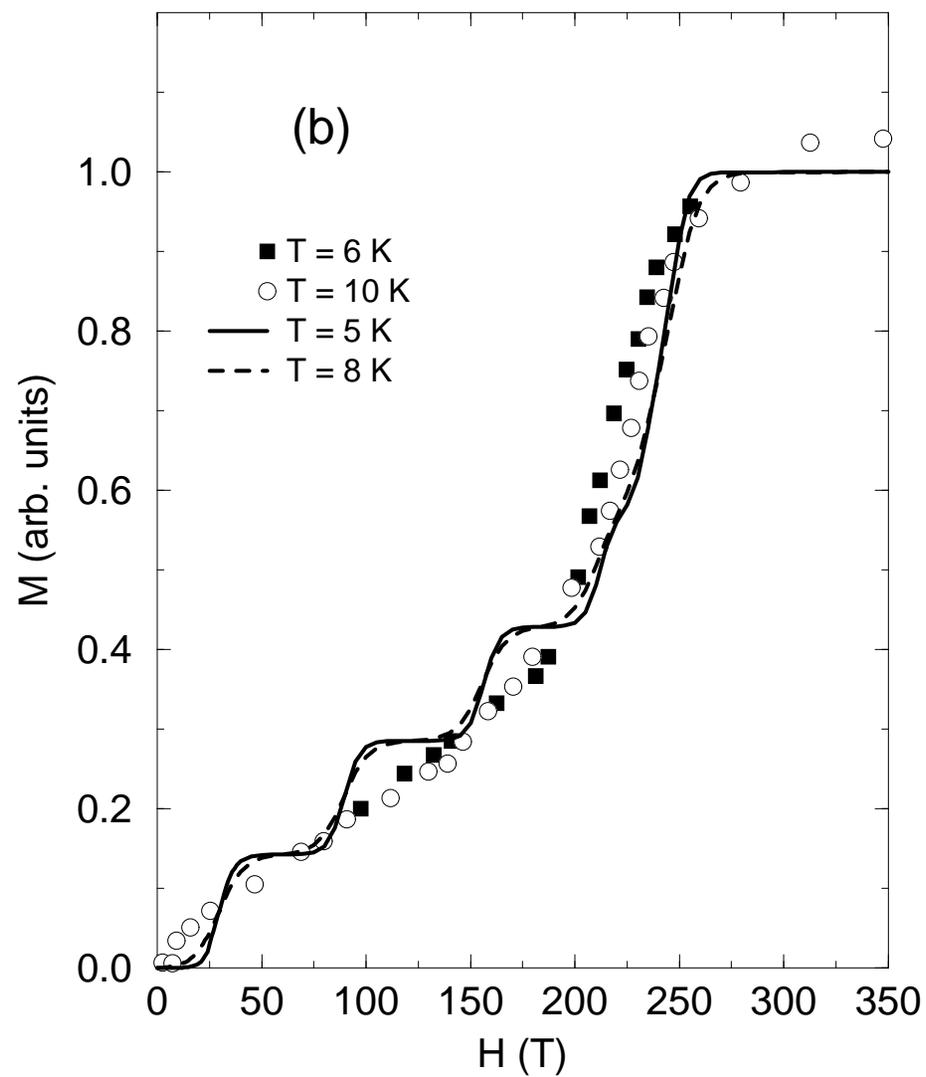
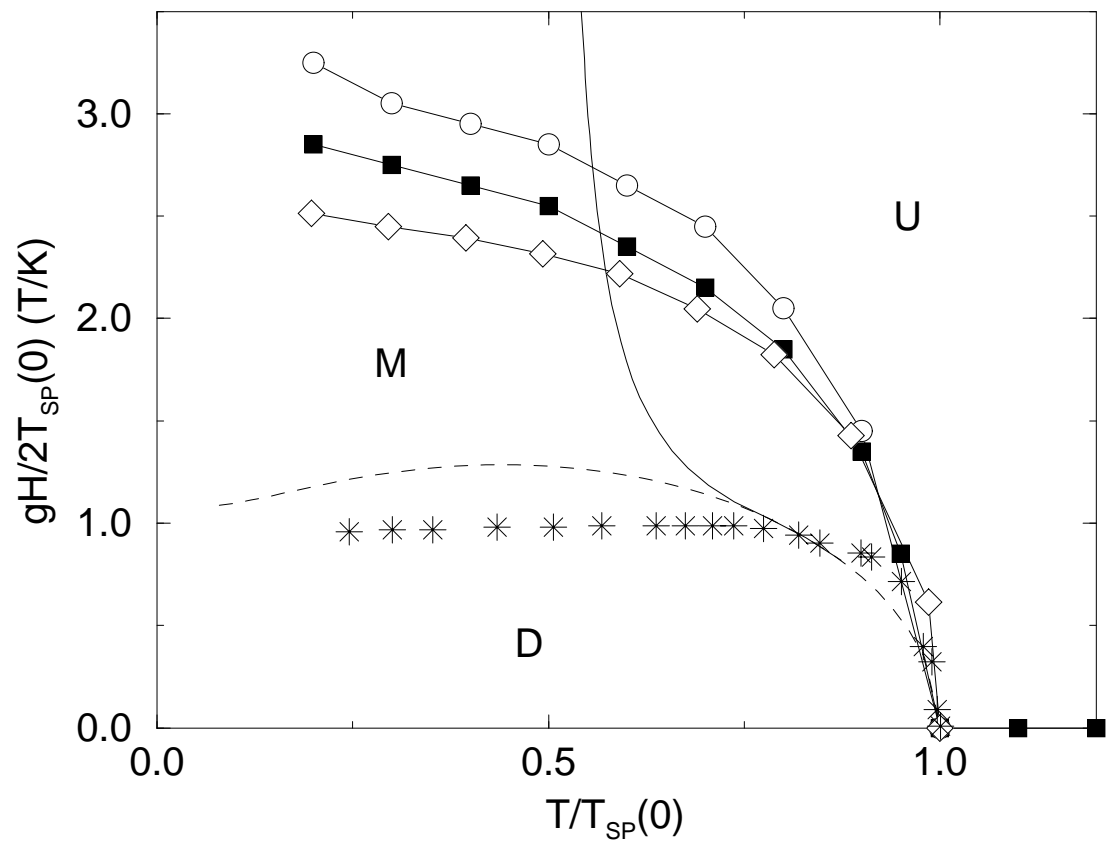
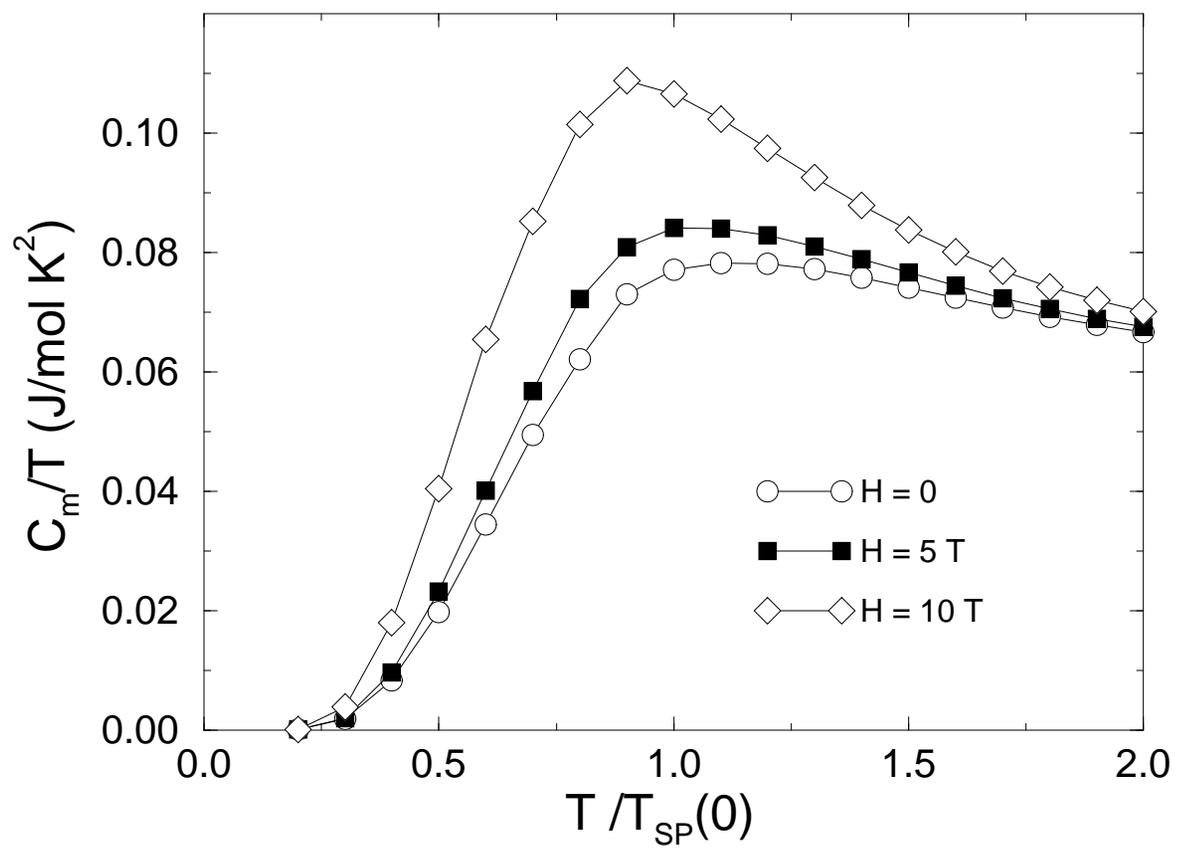
\end{document}